\documentclass{article}
\usepackage[margin=1in]{geometry}
\usepackage{amsmath}
\usepackage{graphicx}

\begin{document}

\title{Definitive Proof of the Classical Multiverse!}
\author{Brian R. La Cour\thanks{Corresponding author: blacour@arlut.utexas.edu} and Noah A. Davis}
\date{Applied Research Laboratories, The University of Texas at Austin}
\maketitle

\begin{abstract}
Recent astonishing experiments with quantum computers have demonstrated unambiguously the existence of a quantum multiverse, where calculations of mind-boggling complexity are effortlessly computed in just a few minutes. Here, we investigate whether a similar computation on a digital computer can demonstrate the existence of a \emph{classical} multiverse. To this end we describe a classical algorithm for efficiently sampling from a $d^n$-dimensional discrete probability distribution representing $n$ digits of $d$ possible values with strong statistical dependence. Although the full distribution for large $n$ quickly becomes intractable, probabilities for given samples can be computed quite efficiently. This allows us to compute exact empirical linear cross-entropy benchmark (XEB) values. Results on a low-end laptop for $d=2$ show excellent agreement with the true XEB for $n \le 30$ and large positive values of the exact empirical XEB for $n \le 1023$ computed over one million samples.  We conclude that classical, as well as quantum, computation occurs in many parallel universes.
\end{abstract}

%%%%%%%%%%%%%%%%%%%%%%%%%%%%%%%%%%%%%%%%%%%%%%%%%%%%%%%

\section{What Up?}

Although quantum computers are still in their early stages of development, they have already demonstrated several amazing and mind boggling phenomena. For example, in 2022, researchers from Google used a mere nine entangled qubits to perform an experimental demonstration of a traversable wormhole \cite{Wormhole}!  Not to be outdone, experimenters from Singapore that same year managed to entangle two superconducting qubits held in high vacuum with a living tardigrade \cite{Tardigrade}!  Of course, Shor's algorithm, implemented on a fault-tolerant quantum computer, will destroy RSA encryption \cite{Shor1997},  but that has already been done using a D-Wave machine \cite{Chao2024}.  It is also a well-established fact, or at least a definite possibility, that quantum computers will solve the climate crisis \cite{ClimateChange,TroyerClimate}.

Notwithstanding these many practical applications, several groups have used quantum devices to perform random sampling in order to demonstrate quantum supremancy or, what in more polite society is now called, \emph{quantum advantage}.  A common example of this is a random circuit experiment, where a random set of one- and two-qubit gates is applied to create a high-dimensional, highly entangled, and generally hard to figure out quantum state.  Measurements of this state in the computational basis produce binary strings that are random realizations from the corresponding classical probability distribution.  The computational challenge is to drawn samples faithfully from the ``true,'' as defined by the set of gates, distribution without everything going to heck because of device imperfections.  Quantum computers can quickly produce samples, but the results aren't great.  Classical computers can compute the distributions exactly, but they take forever.  The ratio of these two time scales is typically what's reported as the quantum advantage.  An excellent summary of the many demonstrations, and refutations, of quantum advantage has been compiled by LaRose \cite{LaRose2024}.  Not included in LaRose's brief history are the recent results from Google's sick new Sycamore chip \cite{Sycamore2024}.  Using 67 qubits and a depth of 32 cycles they were able to demonstrate an astonishing quantum advantage of five minutes to 10 septillion years, leading the Founder and Lead of Google Quantum AI to conclude that this result ``lends credence to the notion that quantum computation occurs in many parallel universes''  \cite{GoogleMultiverse}.

Like many, we were excited by this result and wondered whether a similar demonstration could be performed on a classical digital computer.  After all, the experiment boils down to sampling from an unknown, and practically unknowable, classical probability distribution, albeit one derived from a quantum device.  Could one sample from a classical device with a similarly unknowable probability distribution?  If successful, perhaps that would lend similar credence to the notion of a classical multiverse, where classical computations of such vast complexity occur in many parallel universes.  It should have seemed an impossible task from the start, but we're pretty obtuse and tried it anyway.  Turns out, we could do it!

%%%%%%%%%%%%%%%%%%%%%%%%%%%%%%%%%%%%%%%%%%%%%%%%%%%%%%%

\section{Method to the Madness}

Random circuit experiments use the linear cross-entropy benchmark (XEB) as a measure for gauging the quality of their samples \cite{Gao2024}.  Positive values are good, large positive values are better, and zeros values are just kind of meh.  The XEB between the true probability mass function (pmf) $p$ and a candidate pmf $q$ over the integer values $\{0, 1, \ldots, N-1\}$, where $N$ is a ``pretty big'' integer, is defined to be\cite{XEB2018}
\begin{equation}
\mathrm{XEB} = N \sum_{x=0}^{N-1} p(x) q(x) - 1 \; .
\end{equation}
In the absolutely boring case that $p(x) = 1/N$ is uniform, any $q$, whether it's close to $p$ or not, will yield an XEB of exactly zero, which is rather dull.  If, however, $p$ is nonuniform, then the values of $x$ with high probability will cause the XEB to shoot up like a rocket.  When, in fact, $q = p$ we have what we'll call the ``true XEB.''

Of course, computing all these values is rather tiresome.  Suppose we could faithfully draw $M$ samples $x_1, \ldots, x_M$ from $p$, where $M$ is a sorta large integer but not nearly so large as $N$.  (Fortunately, the XEB is easy to estimate, so $M$ needn't be too big \cite{Rinott2022}.)  We can then take $q$ to be the empirical distribution corresponding to this sample.  Let's furthermore suppose that, by some miracle, would can evaluate $p$ \emph{exactly} at just these values.  The ``exact empirical XEB'' for this sample is then
\begin{equation}
\mathrm{XEB} = \frac{N}{M} \sum_{m=1}^{M} p(x_m) - 1 \; .
\end{equation}

In the quantum case, $p$ is derived from some $N \times N$ unitary matrix $U$ acting on an initial vector $[1, 0, \ldots, 0]^\mathsf{T}$; hence, $p(x) = |U_{x,0}|^2$.  Smartypants mathematicians can show that, if we think of $U$ as a ``Haar-random'' matrix, then $p(x)$ is a random variable that, for large $N$, follows an exponential distribution with a mean value of $1/N$ \cite{PorterThomas1956}.  (Smartypants mathematicians call this a ``Porter-Thomas'' distribution --- whatever.)  In general, though, $p$ looks nothing like a uniform distribution.  This is good because otherwise the XEB would be close to zero, and that wouldn't be any fun at all.  Now, although $U$ consists of a whole lot of elements, in the quantum supremacy --- ahem, quantum advantage --- experiments performed by Google and others, $U$ is generated by a random circuit that only required a ``small'' number of parameters to specify it.  To create our classical multiverse, let's see if we can do something similar --- specify, let's define a huge probability distribution with just a few parameters and draw a bunch of samples from it.

Okay, so on to the multiverse!  Now, it would be stupidly trivial to define $p$ such that $p(x) = 1$ for exactly one value of $x$, thereby guaranteeing that the exact empirical XEB is always $N$.  Let's do something more interesting!  Suppose $N = d^n$ for integers $d \ge 2$ and $n \ge 1$, where $n$ indicates the number of digits with values in $\{0, \ldots, d-1\}$.  (Of course, we could just set $d = 2$ and used binary digits, but one of us (ND) wanted to use polyhedral D\&D dice.  We recommend against using $d = 20$, though, for fear of what creatures may be summoned from the multiverse if we roll a critical failure.)  Let $[s_0, \ldots, s_{n-1}]$ be the big-endian representation of the integer $x \in \{0, \ldots, N-1\}$, where, just to be clear,
\begin{equation}
x = \sum_{i=0}^{n-1} s_i \, d^i \; .
\end{equation}

A general pmf can, duh, be written as a product of conditional probabilities like so
\begin{equation}
p(x) = p(s_0, \ldots, s_{n-1}) = p_0(s_0) \, p_1(s_1|s_0) \cdots p_{n-1}(s_{n-1}|s_0, \ldots, s_{n-2}) \; .
\end{equation}
Our clever idea was to let $w_i = [w_i(j)]_{j=0}^{d-1}$ be a pmf for each $i \in \{0, \ldots, n-1\}$ and define $p$ such that
\begin{equation}
p_i(s_i|s_0, \ldots, s_{i-1}) = w_i(s_0 \oplus \cdots \oplus s_i) \; ,
\end{equation}
where $\oplus$ is addition modulo $d$.  Thus, each $p$ is specified by a mere $nd$ parameters but is defined over $d^n$ values, making it a huge vector.  It's easy to verify that $p$, so defined, is indeed a pmf, so we won't bother showing it.  We decided, somewhat arbitrarily, to draw each $w_i(j)$ randomly from a uniform distribution over $(0,1)$ and then normalized them so that the $d$ values sum to one, thereby making $w_i$ a pmf.  Doing so for all $n$ values of $i$ completely defines $p$.  Generating samples from $p$ is super easy: we simply draw the digits from each successive conditional distribution, which can be done efficiently.  Given $M$ samples, $x_1, \ldots, x_M$, we can also efficiently compute each of the corresponding probabilities, $p(x_1), \ldots, p(x_M)$.  From this, we can compute the exact empirical XEB.  For $M$ sorta large, this should provide a pretty good estimate of the true XEB, the one for which $q = p$.

%%%%%%%%%%%%%%%%%%%%%%%%%%%%%%%%%%%%%%%%%%%%%%%%%%%%%%%

\section{Cool Results}

Being rather strapped for cash, we implemented the algorithm in Matlab on a low-end Dell Precision 3480 laptop with an unimpressive Intel i7-1360P CPU clocking at a mere 2.2 GHz and a pathethic 16 GB of RAM.  For $d = 2$ this limited us to $n \le 30$ for an exact and complete computation of $p$.  Nevertheless, we were about to get some pretty cool results, which we'll now discuss.

\begin{figure}
\begin{center}
\scalebox{0.5}{\includegraphics{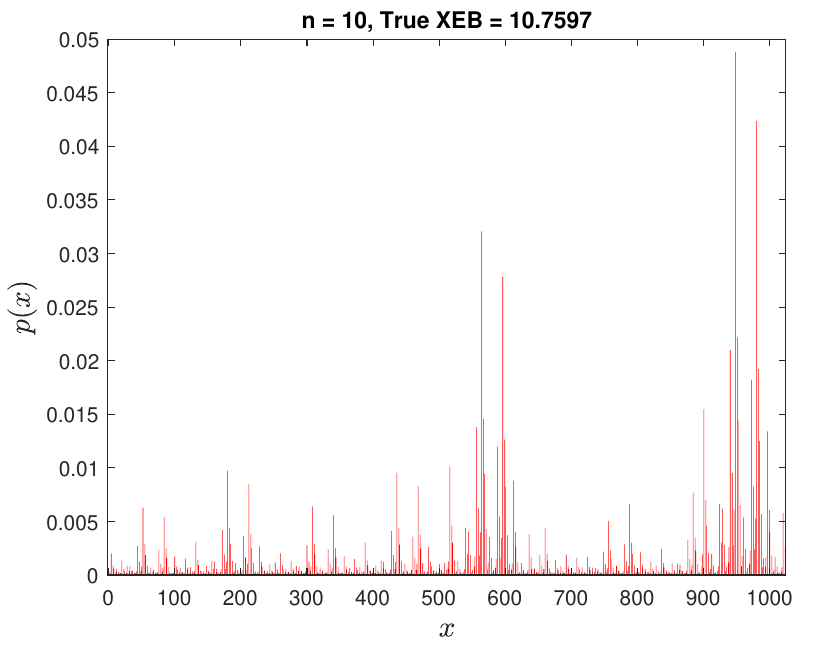}} \scalebox{0.5}{\includegraphics{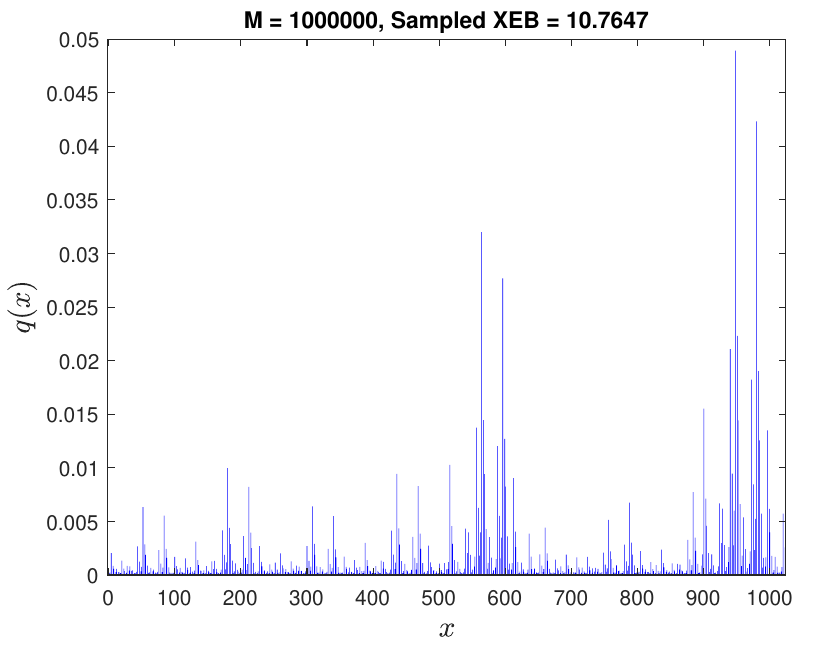}}
\end{center}
\caption{Plot of the exact distribution for $n = 10$ and a random realization of $p$ (red, left) against the empirical distribution computed over a million samples (blue, right).}
\label{fig:pmf10}
\end{figure}

\begin{figure}
\begin{center}
\scalebox{1.0}{\includegraphics{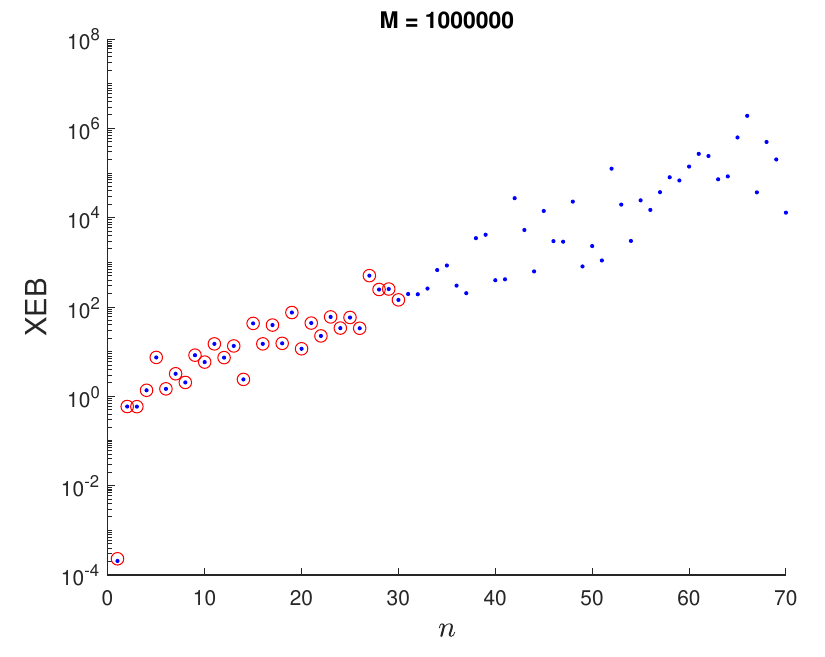}}
\end{center}
\caption{Plot of the exact empirical XEB (blue dots), over one million samples, and the true XEB (red circles) for $d = 2$.  Note that we were only able to compute the true XEB for $n \le 30$.  After that, the machine ran out of memory.}
\label{fig:XEB_vs_n}
\end{figure}

We used $d = 2$, since it was the simplest choice, and generated a different set of parameters for each $n$ to generate a corresponding random realization of $p$.  For $n \le 30$ we were able to compute all values of $p(x)$ and, so, could compute the true XEB.  Beyond $n = 30$, our laptop ran out of memory, so we gave up.  We were also able to compute the exact empirical XEB using a million samples ($M = 10^6$) for each $n$.  An example distribution, for $n = 10$, is shown in Fig.\ \ref{fig:pmf10} and compared against its empirical estimate, which is found to be in quite good agreement.  In Fig.\ \ref{fig:XEB_vs_n} we show a comparison of the exact empirical XEB to the true XEB for $n \le 30$, and we are happy to say that the agreement is excellent.  We were also pleased to see that the XEB values tend to go \emph{up} with increasing $n$, not \emph{down} like in the quantum supremacy experiments.  For fun, we also plotted the exact empirical XEB values for $n$ up to 70, and we note that they follow the same upward trend.  Although we were not able to compute the true XEB for these larger values, the trend seems consistent, so we're pretty darn sure they're right.  Note that the graph is on a semilog scale, so the increase in XEB is actually exponential, which is \emph{really} cool.

\begin{figure}
\begin{center}
\scalebox{1.0}{\includegraphics{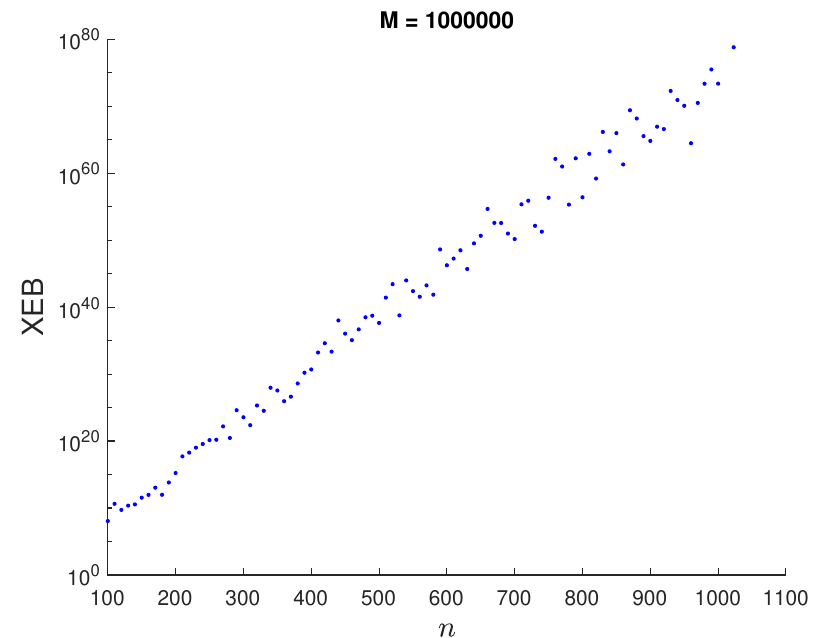}}
\end{center}
\caption{Plot of the exact empirical XEB (blue dots), over one million samples for $d = 2$.  Note that we were only able to compute XEB up $n = 1023$ before it became numerically infinite.}
\label{fig:bigXEB}
\end{figure}

These results are better than anything demonstrated with a quantum computer, but we wanted to see if we could do a little better.  So we continued computing the exact empirical XEB all the way out to $n = 1023$, which was the biggest value we could handle before the XEB became numerically infinite (i.e., could no longer be represented by a double-precision floating point number).  Specifically, in Fig.\ \ref{fig:bigXEB} we show values of the exact empirical XEB for $n \in \{100, 110, \ldots, 1000, 1023\}$, each evaluated over a million samples.  Although we could easily compute $p(x)$ for these million samples, we obviously couldn't compute all $2^n$ values.  We did however estimate, based on our results for $n = 30$, which took about six hours, that to compute the full distribution for $n = 1023$ would take about $10^{300}$ years, though we are doubtful our laptop would hold out that long.  By contrast, to faithfully draw a single sample from this distribution took us only about three microseconds, so this represents a computational advantage of aroud $10^{313}$, which is ginormous!

%%%%%%%%%%%%%%%%%%%%%%%%%%%%%%%%%%%%%%%%%%%%%%%%%%%%%%%

\section{Totally Amazing Conclusion}

We did it!  By tapping into the multiverse we were able to sample from a probability distribution so huge it couldn't possibly fit in just one universe.  The performance of this algorithm is astonishing: It performed a computation in under a millisecond that would take our admittedly lackluster machine $10^{300}$ or a \emph{googol googol googol years}. If you want to write it out, it’s
\begin{quote}
1,000,000,000,000,000,000,000,000,000,000,000,000,000,000,000,000,000,000,000,000,000,000,000,000,
  000,000,000,000,000,000,000,000,000,000,000,000,000,000,000,000,000,000,000,000,000,000,000,000,
  000,000,000,000,000,000,000,000,000,000,000,000,000,000,000,000,000,000,000,000,000,000,000,000,
  000,000,000,000,000,000,000,000,000,000,000,000,000,000,000,000,000,000,000,000,000,000,000,000,
  000,000,000,000 years.
\end{quote}
This mind-boggling number exceeds known timescales in physics and vastly exceeds the age of the universe. It lends credence to the notion that classical, as well as quantum, computation occurs in many parallel universes, in line with the idea that we live in a multiverse, a prediction first made by David Deutsch \cite{Deutsch1997}.

Of course, haters will argue that this is just a cheap trick.  After all, you can't have a multiverse without quantum, and there's clearly no quantum here!  Furthermore, we never used complex numbers, and you clearly cannot have quantum advantage with an imaginary number in there somewhere \cite{Renou2021}.  One might also argue that, although there are strong statistical interdependences baked into the probability distributions, there's no \emph{entanglement}, and you need entanglement for quantum advantage \cite{JozsaLinden2003}.  Of course, we're not looking for quantum advantage, only evidence for a multiverse.  The more high-minded quantum theorists might argue that we have sampled from only \emph{one} probability distribution, whereas a quantum computer creates a quantum state that can be measured in \emph{infinitely many} bases, resulting in infinitely many probability distributions --- indeed, uncountably many!  We admit that this is true but note that each such choice of basis results in just one probability distribution.  Furthermore, any given basis measurement can simply be regarded as a new, slightly modified random circuit measured in the plain old computational basis.  So, there's really no basis for this argument.  We therefore maintain, quite seriously, that our results are as much a demonstration of the classical multiverse and those of quantum computers are a demonstration of the quantum multiverse.

%%%%%%%%%%%%%%%%%%%%%%%%%%%%%%%%%%%%%%%%%%%%%%%%%%%%%%%

\section*{Acknowledgments}

Our sponsor was too embarrased to have their organization affiliated with this paper.  Others we consulted preferred not to be mentioned.

%%%%%%%%%%%%%%%%%%%%%%%%%%%%%%%%%%%%%%%%%%%%%%%%%%%%%%%

% \bibliographystyle{unsrt}
% \bibliography{refs}

\begin{thebibliography}{10}

\bibitem{Wormhole}
D.~Jafferis, A.~Zlokapa, J.~Lykken, D.~Kolchmeyer, S.~Davis, N.~Lauk, H.~Neven,
  and M.~Spiropulu.
\newblock Traversable wormhole dynamics on a quantum processor.
\newblock {\em Nature}, \textbf{612}:51, 2022.

\bibitem{Tardigrade}
K.~Lee, Y.~Tan, L.~Nguyen, R.~Budoyo, K.~Parl, C.~Hufnagel, Y.~Yap,
  N.~M\o{}berg, V.~Vedral, T.~Paterek, and R.~Dumke.
\newblock Entanglement in a qubit-qubit-tardigrade system.
\newblock {\em New Journal of Physics}, \textbf{24}:123024, 2022.

\bibitem{Shor1997}
P.~Shor.
\newblock Polynomial-time algorithms for prime factorization and discrete
  logarithms on a quantum computer.
\newblock {\em SIAM Journal on Scientific Computing}, \textbf{41}:303, 1997.

\bibitem{Chao2024}
W.~Chao et~al.
\newblock Quantum annealing public key cryptographic attack algorithm based on
  {D-Wave} advantage.
\newblock {\em Chinese Journal of Computers}, \textbf{47}:1030, 2024.
\newblock http://cjc.ict.ac.cn/online/onlinepaper/wc-202458160402.pdf.

\bibitem{ClimateChange}
M.~Celsi and L.~Celsi.
\newblock Quantum computing as a game changer on the path towards a net-zero
  economy: A review of the main challenges in the energy domain.
\newblock {\em Energies}, \textbf{17}:1039, 2024.

\bibitem{TroyerClimate}
M.~Troyer.
\newblock State-of-the-art algorithm accelerates path for quantum computers to
  address climate change.
\newblock Published Online 2020-07-30.
\newblock
  https://www.microsoft.com/en-us/research/blog/state-of-the-art-algorithm-accelerates-path-for-quantum-computers-to-address-climate-change/.

\bibitem{LaRose2024}
R.~LaRose.
\newblock A brief history of quantum vs classical computational advantage.
\newblock 2024.

\bibitem{Sycamore2024}
A.~Morvan, B.~Villalonga, X.~Mi, et~al.
\newblock Phase transitions in random circuit sampling.
\newblock {\em Nature}, \textbf{634}:328, 2024.

\bibitem{GoogleMultiverse}
H.~Neven.
\newblock Meet {W}illow, our state-of-the-art quantum chip.
\newblock Published Online 2024-12-09.
\newblock https://blog.google/technology/research/google-willow-quantum-chip/.

\bibitem{Gao2024}
X.~Gao, M.~Kalinowski, C.-N. Chou, M.~Lukin, B.~Barak, and S.~Choi.
\newblock Limitations of linear cross-entropy as a measure for quantum
  advantage.
\newblock {\em PRX Quantum}, \textbf{5}:010334, 2024.

\bibitem{XEB2018}
S.~Boixo, S.~Isakov, V.~Smelyanskiy, R.~Babbush, N.~Ding, Z.~Jiang, M.~Bremner,
  J.~Martinis, and H.~Neven.
\newblock Characterizing quantum supremacy in near-term devices.
\newblock {\em Nature Physics}, \textbf{14}:595, 2018.

\bibitem{Rinott2022}
Y.~Rinott, T.~Shoham, and G.~Kalai.
\newblock Statistical aspects of quantum supremacy demonstration.
\newblock {\em Statistical Science}, \textbf{37}:322, 2022.

\bibitem{PorterThomas1956}
C.~Porter and R.~Thomas.
\newblock Fluctuations of nuclear reaction widths.
\newblock {\em Physical Review}, \textbf{104}:483, 1956.

\bibitem{Deutsch1997}
David Deutsch.
\newblock {\em The Fabric of Reality}.
\newblock Penguin Books, 1997.

\bibitem{Renou2021}
M.-O. Renou, D.~Trillo, M.~Weilenmann, T.~Le, A.~Tavakoli, N.~Gisin,
  A.~Ac\'{i}n, and M.~Navascu\'{e}s.
\newblock Quantum theory based on real numbers can be experimentally falsified.
\newblock {\em Nature}, \textbf{600}:625, 2021.

\bibitem{JozsaLinden2003}
R.~Jozsa and N.~Linden.
\newblock On the role of entanglement in quantum-computational speed-up.
\newblock {\em Proceedings of the Royal Society A: Mathematical, Physical, and
  Engineering Sciences}, \textbf{459}:2011, 2003.

\end{thebibliography}

\end{document}